\documentclass[prl,epsfig,floats,twocolumn,superscriptaddress,amssymb,amsmath,floatfix,showpacs]{revtex4-1}  
\usepackage{graphicx}
\usepackage{bm}  
\usepackage{amssymb}
\usepackage{color}
\usepackage{amscd}

\begin{document}

\title{Direct Measurement of the Free Energy of Aging Hard-Sphere Colloidal Glasses}

\author{Rojman Zargar}
\affiliation{Van der Waals-Zeeman Institute, Institute of Physics (IoP) of the Faculty of Science (FNWI) University of Amsterdam, 1098 XH Amsterdam, the Netherlands}
\author{Bernard Nienhuis}
\affiliation{Institute for Theoretical Physics, Institute of Physics (IoP) of the Faculty of Science (FNWI) University of Amsterdam, 1098 XH Amsterdam, the Netherlands}
\author{Peter Schall}
\affiliation{Van der Waals-Zeeman Institute, Institute of Physics (IoP) of the Faculty of Science (FNWI) University of Amsterdam, 1098 XH Amsterdam, the Netherlands}
\author{Daniel Bonn}
\affiliation{Van der Waals-Zeeman Institute, Institute of Physics (IoP) of the Faculty of Science (FNWI) University of Amsterdam, 1098 XH Amsterdam, the Netherlands}


\begin{abstract}
The nature of the glass transition is one of the most important unsolved problems in condensed matter physics. The difference between glasses and liquids is believed to be caused by very large free energy barriers for particle rearrangements; however so far it has not been possible to confirm this experimentally. We provide the first quantitative determination of the free energy for an aging hard-sphere colloidal glass. The determination of the free energy allows for a number of new insights in the glass transition, notably the quantification of the strong spatial and temporal heterogeneity in the free energy. A study of the local minima of the free energy reveals that the observed variations are directly related to the rearrangements of the particles. Our main finding is that the probability of particle rearrangements shows a power law dependence on the free energy changes associated with the rearrangements, similarly to the Gutenberg-Richter law in seismology.
\end{abstract}
\maketitle

\begin{figure}
\includegraphics[width=0.65\columnwidth]{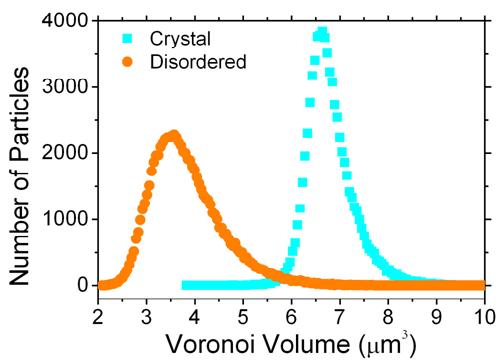}
\caption{Distribution of the Voronoi volumes for the crystal and the glass at $\phi=0.60$.}
\label{Fig1}
\end{figure}

\begin{figure}
\includegraphics[width=1.05\columnwidth]{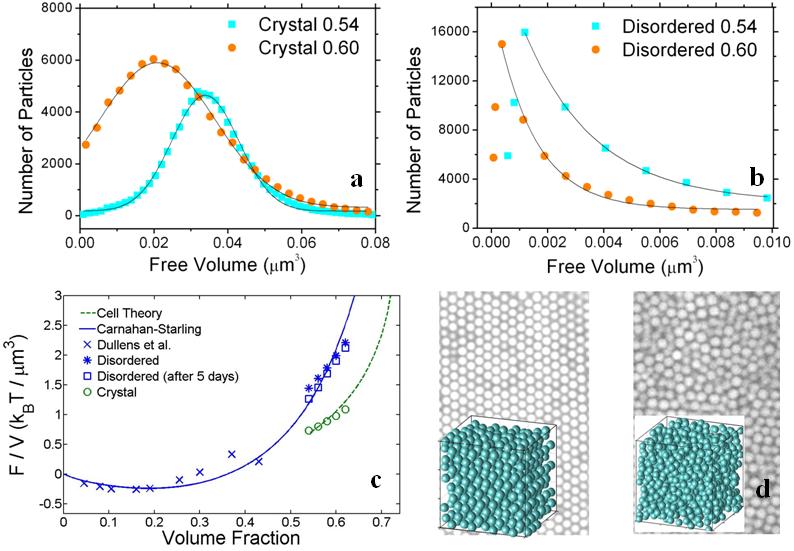}
\caption{(a),(b) Distribution of the free volumes at $\phi=0.54$ and $\phi=0.60$. (a) Crystal; solid curves show Gaussian fit.  (b) Disordered system; solid curves show exponential fit. (c) The free energy density $F/V$ in units of $k_BT/\mu m^3$ as a function of the volume fraction. Circles denote our free energy results for the crystal at volume fractions $0.54$, $0.56$, $0.58$, $0.60$ and $0.62$ that agree with the Hall \cite{Hall} free energy (dashed curve). Stars show the free energy measured for the disordered systems at volume fractions $0.54$, $0.56$, $0.58$, $0.60$ and $0.62$ at the beginning of the measurements while squares indicate the measured free energy after $5$ days. The results agree rather well with the standard Carnahan-Starling \cite{Carnahan} free energy (solid curve). Crosses are the measurements of Dullens \emph{et al.} \cite{Dullens} which are also compared to the Carnahan-Starling result. (d) Typical images of the confocal microscope and a part of the 3D reconstruction of the particle positions. Left: Crystal at $\phi=0.54$, Right: Glass at $\phi=0.60$.}
\label{Fig2}
\end{figure}

\begin{figure}
\centering
\includegraphics[width=1.05\columnwidth]{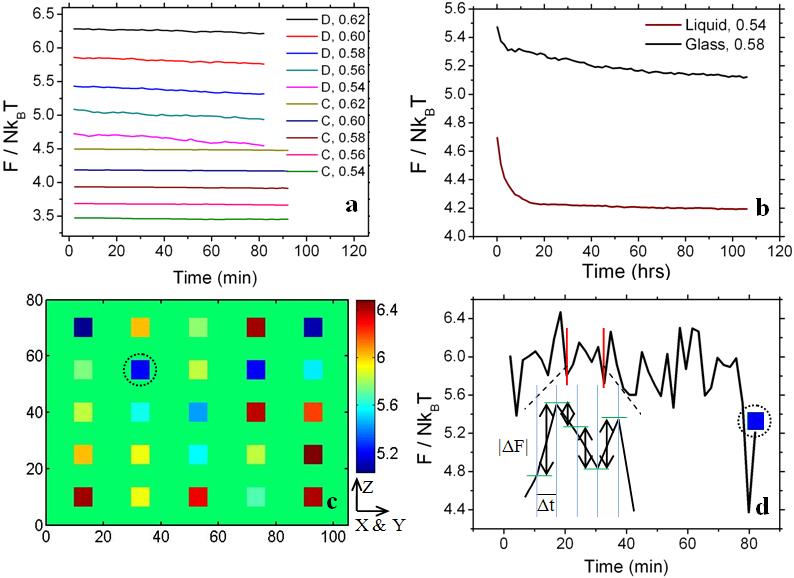}
\caption{(a) Time evolution of the free energy. Free energy per particle is plotted versus time: Crystal (lower part) and disordered system (upper part) at different volume factions. 'D' stands for disordered and 'C' for crystal. (b) Aging of disordered systems. Free energy per particle is plotted versus time for $\phi=0.54$ (brown) and $\phi=0.58$ (black). The noise shows the effect of the local minima of the free energy. Note that the data in (a) corresponds to the very beginning of these curves. (c) Heterogeneity in the free energy distribution in a glass at $\phi=0.60$ at time $t=82 min$ . The color code shows the free energy per particle. The background indicates the free energy calculated for the whole system. (d) The free energy per particle versus time in one of the boxes considered in (c).}
\label{Fig3}
\end{figure}

The simplest system that has become a paradigm for molecular glasses is a colloidal hard-sphere system in which spheres interact only through an infinite repulsion on contact \cite{Goetze}. The phase behavior of such hard-sphere systems is uniquely governed by the volume fraction $\phi$ \cite{Pusey}: There is a liquid phase for $\phi<0.494$, a crystalline phase for $\phi>0.545$ and coexistence between the two for $0.494<\phi<0.545$. If crystallization is avoided, for $0.545<\phi<0.58$ the system is a supercooled liquid whereas for $\phi>0.58$ the dynamics of the system becomes so slow that it is a glass \cite{Debenedetti, Kegel,Weeks_heterogeneity,Kawasaki,Sara_heterogeneity}.
\\ Because of the absence of interaction energies, the phase behavior of hard-sphere systems is dictated only by entropic contributions, implying that the statistical geometry provides a direct route to study the thermodynamics of the system \cite{Speedy}. The usual way to do this is to determine the available volume, i.e. the volume available to insert an additional sphere into the system; there are exact relations that relate the available volume to the thermodynamics of the system \cite{Sastry,Dullens}. This method was successfully applied to colloidal liquids, but is unfortunately not applicable to glasses, because the cavities become extremely small at high densities and so the determination of the available volume becomes prohibitively difficult \cite{Dullens}.
\\ An alternative method relies on the use of the related free volume \cite{Bouehler}, the volume over which the center of sphere can move when all other particles are held fixed. Unlike the available volume, the free volume is a local quantity that can be obtained for each individual spherical particle. Since the free volume is typically much larger, it can be measured with much greater accuracy.
\\ In this Letter, we apply the free volume method to measure the free energy of a colloidal glass. In this approach the system is divided into $N$ cells constructed around the particle positions; choosing the cells so that they are singly occupied, the free energy of a hard-sphere system for a given configuration of the particles can be expressed directly in terms of only the free volume as \cite{Coniglio}:
\begin{equation}
F \simeq -k_BT \sum_{i=1}^N \ln (\frac{v_{f_i}}{\lambda^3}),
\label{freeenergy1}
\end{equation}
in which $v_{f_i}$ is the free volume for particle '$i$' in its cell, and $\lambda=(h^2/2\pi mk_BT)^{1/2}$ is the thermal wavelength; this form supposes there are no interactions other than hard-sphere.
\\ We determine $v_f$ experimentally in a colloidal hard-sphere system;
we use sterically stabilized fluorescent poly-methylmethacrylate (PMMA) particles suspended in a refractive index and density matched mixture of cis-decalin and cycloheptyl bromide (CHB) which are the best model system for hard-spheres \cite{Royall}. Our particles have a diameter $\sigma=1.5 \mu m$ and polydispersity $\simeq 7\%$ to prevent crystallization. The crystal is obtained in a second system with $\sigma=1.7 \mu m$ and polydispersity $\simeq 4\%$. In both systems an organic salt,  TBAB (tetrabutylammoniumbromide), is used to screen any possible residual charges. Suspensions at volume fractions between $0.54$ and $0.62$ are prepared by diluting sediments that are centrifuged to random close packing ($\phi_{rcp}\simeq 0.64$); for a discussion on the accuracy of this procedure, see Ref. \cite{Poon}. In order to verify the hard-sphere nature of the colloids the crystallization density, a very sensitive measure for deviations from the hard-sphere behavior, is measured: The results agree to that of the true hard-spheres within a fraction of a percent. To initialize the glassy samples we insert a small stirring bar into the chamber of each sample to stir and have a reproducible initial state for the sample at the start of the experiments. The initial time $t=0$ is defined as the beginning of the measurements that we start $20$ minutes after ending the stirring.
\\ Using a fast confocal microscope (Zeiss LSM live) we acquire 3D images of the system in a $105\mu m \times 105\mu m \times 80\mu m$ volume to find the positions of $\simeq2.5\times10^5$ individual particles in three dimensions (Fig. \ref{Fig2}(d)). We do so for both disordered (liquid, glass) and ordered (crystalline) systems. Using the positions of particles, a Voronoi tessellation is generated \cite{Voronoi}: A Voronoi tessellation divides the space into distinct, non-overlapping convex polyhedra which surround each point. The Voronoi volume is given by the polyhedron around the center of a sphere '$i$' consisting of the region of space closer to '$i$' than the center of all the other spheres. We find that the distribution of the Voronoi volumes for the particles in the crystal is much sharper than in the disordered system (Fig. \ref{Fig1}) indicating the same Voronoi volume for the particles in the crystal. In addition, the measured distribution for the Voronoi volumes agrees very well with simulations \cite{Starr,Kumar}.
\\ The free volume however, is defined as the volume over which the center of sphere can move when all other particles are fixed. Using the position of particles, we can then calculate the free volume for each particle which is assumed to be confined to its Voronoi cell. Following Ref. \cite{Coniglio} and consistent with Ref. \cite{Hill,Wood}, the free volume associated with a Voronoi cell is defined as the volume of a smaller cell generated from the Voronoi cell by moving the faces normally inside of a distance $\sigma /2$, where $\sigma$ is the particle diameter. It should be noted that the assumption of considering that the particles are confined in cells is valid only at high densities where the particles are so densely packed that no rearrangements happen on the time scale of obtaining the configuration.
\\ The observed distribution for the free volumes in the crystals is consistent with a Gaussian distribution (Fig. \ref{Fig2}(a)). For the glass however, the observed distribution for the free volumes shows an exponential tail (Fig. \ref{Fig2}(b)) in line with random nature of the glassy state. For the same volume fraction, the average free volume for particles in the glass is much smaller than that in the crystal (Fig. \ref{Fig2}(a),(b)), leading to a larger free energy for the glass; this is of course expected, because the crystal is the thermodynamically stable phase.
\\ The colloidal crystal is used as a reference system for our measurements since there are other very accurate methods to determine the free energy for ordered systems \cite{Hall,Rutgers}. We find that the measured free energies for the hard-sphere crystal are in excellent agreement with the Hall equation of state (dashed curve in Fig. \ref{Fig2}c). On the other hand, the free energy for the disordered system for volume fractions up to the glass transition are found to be in good agreement with the standard Carnahan-Starling equation of state which is known to be very accurate for the liquid phase of the hard-sphere systems \cite{Carnahan}. The free energy results of Dullens \emph{et al.} \cite{Dullens}, who were only able to measure the free energy for $\phi<0.40$ because they were using the available volume method, also agree with the Carnahan-Starling results.
\\ This method then allows, for the first time, to investigate the free energy and its variations as function of both space and time for a glass. In Fig. \ref{Fig3}(a) we show that, while the free energy of the crystal is constant as expected for an equilibrium system, it decreases significantly over a long time in disordered systems, signaling aging of the system \cite{Daniel_rejuvenation,Sara_flucdissipation}. This is what is commonly assumed for the time evolution of non-equilibrium glassy systems, but we are the first to observe the temporal free energy decrease experimentally. For hard-sphere colloids it is well known that aging can be observed on similar time scales in e.g. the correlation function of the particle displacements \cite{Courtland,Cianci}.
\\To check the effect of the particles size distribution on the calculated free energy, we add size polydispersity into the crystal as well as the glass artificially: A polydispersity of $7\%$ distributed randomly in the glass decreases the free energy by $6\%$ while a polydispersity of $4\%$ randomly distributed in the crystal increases the free energy by an amount of $4\%$. This is much smaller than the typical fluctuations in the free energy and the difference between ordered and disordered systems that we consider here.
\\ To see whether there is a clear signature of glassy behavior, we measure the free energy for several days for a disordered system with $\phi=0.54$ and for $\phi=0.58$, which are commonly believed to be a supercooled liquid and a glass (Fig. \ref{Fig3}(b)) \cite{Debenedetti}. In the supercooled liquid the free energy saturates to a horizontal plateau. For the glassy system at $\phi=0.58$, after the quench into the glassy phase, the free energy evolves very slowly and indeed the glass never equilibrates on the experimental time scale (Fig. \ref{Fig3}(b)), in perfect agreement with the usual assertion that the system becomes glassy around $\phi=0.58$ \cite{Debenedetti}. We also observe rather large fluctuations in the time dependence of free energy suggesting that every now and then the system becomes trapped in local free energy minima (Fig. \ref{Fig3}(d)), as is also expected for a glass \cite{Dasgupta, Parisi}.
\\ To look at the aging in more detail, we investigate small subsystems, so that the fluctuations are more pronounced. We calculate the free energy for many small boxes, each containing around $60$ particles, in the colloidal glass at $\phi=0.60$ (Fig. \ref{Fig3}(c)). The free energy of the glass is found to be very heterogeneous: The free energy varies strongly with the position of the box, e.g. the free energy may change about $20\%$, or more than $1k_BT$ per particle from one region to another at any given time.
\begin{figure}
\includegraphics[width=0.70\columnwidth]{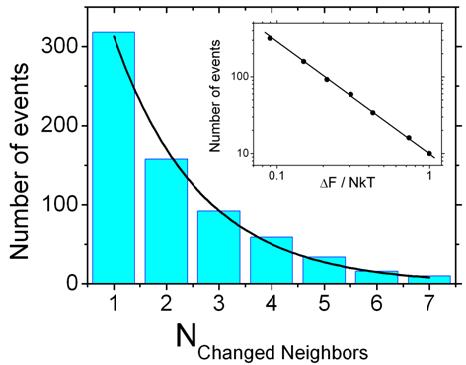}
\caption{Probability distribution of neighbor changing events. The number of events decreases exponentially with the number of changed neighbors in the event. The solid curve is an exponential fit. Inset: The number of events over a given time period varies as a power law with the free energy change associated with the events.}
\label{Fig4}
\end{figure}
\\ We are now in the position to address the central question about glassy dynamics, namely how the aging of the free energy is related to the rearrangements of particles.
To investigate this, we computed the number of changed nearest neighbors for each pairs, located in regions that contain around $200$ particles, from the Voronoi decomposition \cite{Vijay}. Comparing the neighbors of all the particles after each time step allows us to detect rearrangements as events for which a certain number of neighbors change between two time steps (Fig. \ref{Fig4}). We find that the probability of occurrence of a neighbor changing event decreases exponentially with the magnitude of the event, i.e. the number of changed neighbors. If we take $\Delta F$ to be the absolute value of the free energy difference between two consecutive snapshots (see Fig. \ref{Fig3}(d)) averaged over events with the same number of changed neighbor pairs, we find that the number of events obeys a power law scaling in terms of the free energy changes with a power $\simeq -1.2\pm 0.15$ (Fig. \ref{Fig4}). Surprisingly this is analogous to the Gutenberg-Richter law \cite{Gutenberg} in seismology where the probability for large earthquakes to happen decreases exponentially with the magnitude of the earthquakes, the latter being logarithmic in terms of the released energy; in this case number of events is a power law in terms of the released energy with a power $\simeq -1\pm 0.4$ \cite{Utsu}.
\\ In summary, we present the first measurements of the free energy of a glassy system. The concept of the free energy landscape and its deep minima is central in our current theoretical understanding of glassy systems. However, a direct measurement was so far not possible but has been achieved here and reveals the strong heterogeneity of the free energy and its link to the slow structural relaxation that is the hallmark of glassy behavior.
\\ We would like to thank the Stichting voor Fundamenteel Onderzoek der Materie (FOM) and Shell for the financial support of the present research work.


\begin{thebibliography}{0}%
\makeatletter
\providecommand \@ifxundefined [1]{%
 \@ifx{#1\undefined}
}%
\providecommand \@ifnum [1]{%
 \ifnum #1\expandafter \@firstoftwo
 \else \expandafter \@secondoftwo
 \fi
}%
\providecommand \@ifx [1]{%
 \ifx #1\expandafter \@firstoftwo
 \else \expandafter \@secondoftwo
 \fi
}%
\providecommand \natexlab [1]{#1}%
\providecommand \enquote  [1]{``#1''}%
\providecommand \bibnamefont  [1]{#1}%
\providecommand \bibfnamefont [1]{#1}%
\providecommand \citenamefont [1]{#1}%
\providecommand \href@noop [0]{\@secondoftwo}%
\providecommand \href [0]{\begingroup \@sanitize@url \@href}%
\providecommand \@href[1]{\@@startlink{#1}\@@href}%
\providecommand \@@href[1]{\endgroup#1\@@endlink}%
\providecommand \@sanitize@url [0]{\catcode `\\12\catcode `\$12\catcode
  `\&12\catcode `\#12\catcode `\^12\catcode `\_12\catcode `\%12\relax}%
\providecommand \@@startlink[1]{}%
\providecommand \@@endlink[0]{}%
\providecommand \url  [0]{\begingroup\@sanitize@url \@url }%
\providecommand \@url [1]{\endgroup\@href {#1}{\urlprefix }}%
\providecommand \urlprefix  [0]{URL }%
\providecommand \Eprint [0]{\href }%
\providecommand \doibase [0]{http://dx.doi.org/}%
\providecommand \selectlanguage [0]{\@gobble}%
\providecommand \bibinfo  [0]{\@secondoftwo}%
\providecommand \bibfield  [0]{\@secondoftwo}%
\providecommand \translation [1]{[#1]}%
\providecommand \BibitemOpen [0]{}%
\providecommand \bibitemStop [0]{}%
\providecommand \bibitemNoStop [0]{.\EOS\space}%
\providecommand \EOS [0]{\spacefactor3000\relax}%
\providecommand \BibitemShut  [1]{\csname bibitem#1\endcsname}%
\let\auto@bib@innerbib\@empty
\end{thebibliography}%


\begin{thebibliography}{9}

\bibitem{Goetze}
W. Goetze,  \emph{Liquids, Freezing, and Glass Transition}, Part I, edited by J. P. Hansen, D. Levesque, and J. Zinn-Justin (North-Holland, Amsterdam, 1991).

\bibitem{Pusey}
P. N. Pusey and W. van Megen, Nature \textbf{320}, 340 (1986).

\bibitem{Debenedetti}
P. G. Debenedetti and F. H. Stillinger, Nature \textbf{410}, 259 (2001).

\bibitem{Kegel}
W. K. Kegel and A. van Blaaderen, Science \textbf{287}, 290 (2000).

\bibitem{Weeks_heterogeneity}
E. R. Weeks, J. C. Crocker, A. C. Levitt, A. Schofield, and D. A. Weitz, Science \textbf{287}, 627 (2000).

\bibitem{Kawasaki}
T. Kawasaki and H. Tanaka, J. Phys.: Condens. Matter \textbf{22}, 232102 (2010).

\bibitem{Sara_heterogeneity}
S. Jabbari-Farouji, R. Zargar, G. H. Wegdam, and D. Bonn, Soft Matter \textbf{8}, 5507 (2012).

\bibitem{Speedy}
R. J. Speedy, J. Chem. Soc. Faraday Trans. 2 \textbf{76}, 693 (1980).

\bibitem{Sastry}
S. Sastry, T. M. Truskett, P. G. Debendetti, S. Torquato, and F. H. Stillinger, Molecular Phys. \textbf{95}, 289 (1998).

\bibitem{Dullens}
R. P. A. Dullens, D. G. A. L. Aarts, and W. K.  Kegel, Proc. Natl. Acad. Sci. U.S.A. \textbf{103}, 529 (2006).

\bibitem{Bouehler}
R. J. Bouehler, R. H. Wentorf, J. O. Hirsci-Welder, and C. F. Courtiss, J. Chem. Phys. \textbf{19}, 61 (1951).

\bibitem{Royall}
C. P. Royall, W. C. K. Poon, and E. R. Weeks, Soft Matter \textbf{9}, 17 (2013).

\bibitem{Poon}
W. C. K. Poon, E. R. Weeks, and C. P. Royall, Soft Matter \textbf{8}, 21 (2012).

\bibitem{Coniglio}
T. Aste and A. Coniglio, Europhys. Lett. \textbf{67}, 165 (2004).

\bibitem{Voronoi}
G. F. Voronoi, Journal für die reine und angewandte Mathematik \textbf{134}, 198 (1908).

\bibitem{Starr}
F. W. Starr, S. Sastry, J. F. Douglas, and S. C. Glotzer, Phys. Rev. Lett. \textbf{89}, 125501 (2002).

\bibitem{Kumar}
V. S. Kumar and V. Kumaran, J. Chem. Phys. \textbf{123}, 114501 (2005).

\bibitem{Hill}
T. Hill, \emph{Statistical Mechanics} (Mc Graw-Hill, New York) 1956, Chap. 8.

\bibitem{Wood}
W. W. Wood, J. Chem. Phys. \textbf{20}, 1134 (1952).

\bibitem{Hall}
K. R. Hall, J. Chem. Phys. \textbf{57}, 2252 (1972).

\bibitem{Rutgers}
M. A. Rutgers, J. H. Dunsmuir, J. -Z. Xue, W. B. Russel, and P. M. Chaikin, Phys. Rev. E \textbf{53}, 5043 (1996).

\bibitem{Carnahan}
N. F. Carnahan and K. E. Starling, J. Chem. Phys. \textbf{51}, 635 (1969).

\bibitem{Daniel_rejuvenation}
D. Bonn, S. Tanse, B. Abou, and H. Tanaka, J. Meunier, Phys. Rev. Lett. \textbf{89}, 015701 (2002).

\bibitem{Sara_flucdissipation}
S. Jabbari-Farouji, D.  Mizuno, M. Atakhorrami, F. C. MacKintosh, C. F. Schmidt, E. Eiser, G. H. Wegdam, and D. Bonn, Phys. Rev. Lett. \textbf{98}, 108302 (2007).

\bibitem{Courtland}
R. E. Courtland and E. R. Weeks, J. Phys. Cond. Matter \textbf{15}, S359 (2003).

\bibitem{Cianci}
G. C. Cianci, R. E. Courtland, and E. R. Weeks, Solid State Communications \textbf{139}, 599 (2006).

\bibitem{Dasgupta}
C. Dasgupta and O. T. Valls, Phys. Rev. E \textbf{59}, 3123 (1999).

\bibitem{Parisi}
G. Parisi and F. Zamponi, Rev. Modern Phys. \textbf{82}, 789 (2010).

\bibitem{Vijay}
V. Chikkadi, G. H. Wegdam, D. Bonn, B. Nienhuis, and P. Schall, Phys. Rev. Lett. \textbf{107}, 198303 (2011).

\bibitem{Gutenberg}
B. Gutenberg and C. F. Richter, Geol. Soc. Am., Spec. Pap. \textbf{34}, 1 (1941).

\bibitem{Utsu}
T. Utsu, Pure Appl. Geophys. \textbf{155}, 509 (1999).






\end{thebibliography}
\end{document}